# Experiences of WFPC-2 photometry and PSF modelling

Nial Tanvir, David Robinson, Ted von Hippel

Institute of Astronomy
Madingley Road
Cambridge.

The wide-field chips in WFPC-2 have $\sim 0.0996$ arcsec pixels, which is larger than the FWHM of the point spread function (PSF). This poor sampling of WFPC-2 images, means that simple stellar aperture photometry is competitive with profile fitting for moderately crowded fields. One of the problems which must be addressed with profile fitting photometry is that of allowing for variations of the PSF as a function of position on the image. The equivalent problem for aperture photometry is obtaining aperture corrections as a function of position. The aperture correction must be added to each aperture magnitude in order to account for the flux from the star falling outside the aperture. If the aperture used is large, and with WFPC-2 this means about 1 arcsec in diameter, then the aperture corrections will be both small and fairly constant. However, to obtain better signal-to-noise especially in crowded fields, and hence take full advantage of the superb resolution of HST, it is desirable to use smaller apertures.

### Photometry of NGC2477

We have used $V$ and $I$-band WFPC-2 images (filters F555W and F814W) of the open cluster NGC2477 to explore the issue of aperture corrections. The data were acquired to investigate the white dwarf cooling sequence and the low mass end of the main sequence in this cluster (von Hippel etal., in preparation). However, the large number of bright, but still reasonably uncrowded, stars makes these images particularly suitable for experiments with photometry.

We used the PHOT routine within the IRAF implementation of DAOPHOT-2 (Stetson 1987) to obtain aperture photometry in 3 and 10 pixel diameter apertures, corresponding to 0.3 and 1 arcsec. To ensure good statistics, only stars with a $S/N > 30$, as determined by PHOT, were used. First we plotted the aperture corrections as a function of centroid position within a pixel (figure 1). Since the PHOT routine computes magnitudes by fractional pixel integration of counts within the circular aperture, some effect of pixelation would be expected. This figure shows that any systematic variation of aperture correction with position of stellar centroid within a pixel is $\leq$ 3% for each band, even for such a small aperture. Further experiments are required to assess the significance of the correlation which is seen.

We then proceeded to plot the aperture corrections binned as functions of various geometric parameters. The greatest correlation was seen when the corrections were plotted against the radial distance of the star from the centre of each CCD. This effect (figure 2) seems roughly linear and is similar in both $V$ and $I$. It is in the sense that stars in the centres of the chips are somewhat sharper than those near the edge, amounting to a difference of 0.07 mags in the aperture correction. We note that because the effect is in both bands, at least the derived colours should still be reasonably accurate. This is important for studies which rely on field stars to estimate reddening.

### TinyTim simulations

To investigate this further, we constructed grids of PSFs using the TinyTim v.4.0 image modelling software (Krist 1994). Each simulation consisted of 144 PSFs arranged in a $12 \times 12$ grid at 64 pixel separation. Grids were created for each of the three wide-field CCDs in both $V$ and $I$. We chose to create the PSFs with no simulated "jitter" effect, and for stars of intermediate colour ($B - V = 0.619$). Changing the colour to $B - V = -0.155$ only made a 0.2% effect.

The results of these simulations are presented in graphical form in figure 3. Again we have calculated aperture corrections using PHOT, although this time the "data" has essentially infinite signal-to-noise. It is clear that the sharper PSFs (ie. the smaller aperture corrections) are also generally found in the chip centres in these simulations. However, the apparent variations as a function of row and column, which are largely due to the tilt of each CCD to the focal plane, are rather higher than we see in the real data. Furthermore, TinyTim also predicts a variation from CCD to CCD in the sense that WF2 should have the sharpest PSFs on average and WF3 the broadest. The NGC2477 data do show this effect, although at a lower level.

Using the TinyTim maps to correct the NGC2477 aperture magnitudes (figure 4) shows that the variation of aperture correction as a function of radial position is considerably reduced, although is still present at a lower $\sim 3\%$ level. Not surprisingly the TinyTim simulations on average overcorrect the real data, particularly in the $V$. In other words, the real data are slightly lower resolution than the simulations, which we expect due to factors such as spacecraft "jitter" (Bely 1994), and focus changes produced by OTA "breathing" (Hasan and Bely 1994) and "desorption" (the HST focus changes slowly due to shrinkage of the telescope structure, and is readjusted when this reaches a $5\mu m$ offset.). This suggests that the "average" aperture correction should be obtained directly off the frame, even if some scheme, such as the one outlined here, is used to gauge the position dependent variations. The residual correlation with radius appears real, and might also be a consequence of an off nominal focus position. This is further suggested by the fact that the in correcting the data, we actually introduce a small correlation with column number, particularly in the case of WF3.

The aperture correction from 10 pixels (diameter) to total is given in the WFPC-2 IDT report (Holtzman etal. 1994) as 0.09 mags for $V$ and 0.08 mags for $I$. We also confirm these numbers from a small number of bright isolated stars in the NGC2477 data. The Tinytim simulations produce similar figures out to a 30 pixel diameter aperture, and the variation with position of the 10 to 30 pixel correction is only of order 0.5% ($1\sigma$). Whilst the simulations also predict a few percent of flux at even larger apertures, this is irrelevant providing that standard star calibration is performed consistently at the smaller apertures.

**What next?**

We are conducting further experiments using other WFPC-2 images in order to see how stable these effects are with time. It already appears that the average aperture correction over the whole image varies between data-sets, again possibly due to the effects of jitter, breathing and longer term focus changes. We are hopeful that by modifying the focus parameters within TinyTim we can obtain simulations which much more accurately match the data. Ultimately profile fitting photometry and image deconvolution also need to account for PSF variation, and in many cases it will be difficult to characterise this from the target frames themselves given that it requires a good sample of bright isolated stars on each CCD. Thus, for these techniques too, the ability to generate realistic model PSFs will be extremely valuable.

**Figure Captions.**

Figure 1.

Plot of aperture correction as a function of the centroid position of the star within a pixel. The data for all CCDs are binned and a clipped median is found. The error bars represent the interquartile range of the clipped data, divided by the square root of the number of data values within the bin.

Figure 2.

The aperture correction (between apertures of 3 and 10 pixel diameter) is shown to correlate with radius from the centre of each chip. A similar variation is seen in both $V$ and $I$.

Figure 3.

These maps show graphically the difference in aperture correction (3 to 10 pixels diameter) as a function of position which is predicted by TinyTim. Plots are shown for the three wide-field CCDs on WFPC-2 in each of $V$ and $I$. Contours are spaced at 0.01 magnitude intervals. Variations as high as 0.08 magnitudes are seen across individual chips, and some variation is also predicted between the chips.

Figure 4.

The radial dependence of aperture correction is reduced when the TinyTim predictions are accounted for. The residual difference between the data and the models could be explained by small focus changes which the HST experiences.